# Skitter: A Distributed Stream Processing Framework with Pluggable Distribution Strategies

Mathijs Saey[a], Joeri De Koster[a], and Wolfgang De Meuter[a]

a   Software Languages Lab, Vrije Universiteit Brussel, Belgium

**Abstract**

**Context**  Distributed Stream Processing Frameworks (DSPFs) are popular tools for expressing real-time Big Data applications that have to handle enormous volumes of data in real time. These frameworks distribute their applications over a cluster in order to scale horizontally along with the amount of incoming data.

**Inquiry**  Crucial for the performance of such applications is the *distribution strategy* that is used to partition data and computations over the cluster nodes. In some DSPFs, like Apache Spark or Flink, the distribution strategy is hardwired into the framework which can lead to inefficient applications. The other end of the spectrum is offered by Apache Storm, which offers a low-level model wherein programmers can implement their own distribution strategies on a per-application basis to improve efficiency. However, this model conflates distribution and data processing logic, making it difficult to modify either. As a consequence, today's cluster application developers either have to accept the built-in distribution strategies of a high-level framework or accept the complexity of expressing a distribution strategy in Storm's low-level model.

**Approach**  We propose a novel programming model wherein data processing operations and their distribution strategies are decoupled from one another and where new strategies can be created in a modular fashion.

**Knowledge**  The introduced language abstractions cleanly separate the data processing and distribution logic of a stream processing application. This enables the expression of stream processing applications in a high-level framework while still retaining the flexibility offered by Storm's low-level model.

**Grounding**  We implement our programming model as a domain-specific language, called Skitter, and use it to evaluate our approach. Our evaluation shows that Skitter enables the implementation of existing distribution strategies from the state of the art in a modular fashion. Our performance evaluation shows that the strategies implemented in Skitter exhibit the expected performance characteristics and that applications written in Skitter obtain throughput rates in the same order of magnitude as Storm.

**Importance**  Our work enables developers to select the most performant distribution strategy for each operation in their application, while still retaining the programming model offered by high-level frameworks.

# The Art, Science, and Engineering of Programming

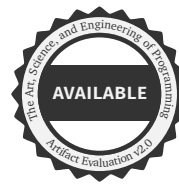
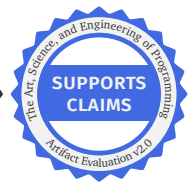

**Skitter: A DSPF with Pluggable Distribution Strategies**

# 1 Introduction

Over the past decade, *Distributed Stream Processing Frameworks* (DSPFs) such as Spark Streaming [3, 27], Flink [7] and Storm [23], emerged as important tools to create scalable data processing applications that respond to incoming streams of "Big Data" in real time [2]. These frameworks gained traction thanks to their ability to distribute these applications over multiple machines connected in a cluster configuration, enabling them to scale with the amount of computational resources.

In such frameworks, applications are expressed by transforming data streams through the use of various *operations* which may, in turn, emit new data streams, such as visually exemplified in Figure 1. The DSPF can then transparently distribute these applications over the cluster computer. When distributing an application, a DSPF is responsible for deciding how the state belonging to each operation is distributed over the cluster, for determining the cluster node on which the computations of each operation are executed, and for communicating between these nodes. We call the logic which governs these decisions for an operation the *distribution strategy* of the operation. Since distribution strategies determine how the work for the operations in an application is divided over the cluster nodes and, since communication between cluster nodes is a costly operation [14], distribution strategies have a major impact on the performance characteristics of a stream processing application [12, 24, 29].

Many modern DSPFs, including Apache Spark and Flink, impose a programming model wherein applications are expressed through a predefined set of operations (such as map, join or reduce), each of which is associated with a built-in distribution strategy. This fixed dependency between operation and distribution strategy is problematic, as the chosen distribution strategy may be a poor fit for the operation or application, and possibly degrade its performance. An example of this can be found in the join operation of Apache Flink, the performance of which is degraded when faced with heavily skewed data [11]. The alternative to this model is offered by Apache Storm, which offers a lower-level programming model that is flexible enough to allow application developers to control precisely how their application is distributed over a cluster. Unfortunately, this flexibility comes at the cost of substantial additional accidental complexity. Moreover, Storm offers no abstractions for keeping the distribution strategy separate from the application code. Instead, a distribution strategy is specified by modifying several distinct locations in the application code. Nevertheless, Storm is the current "gold standard" for expressing distribution strategies, as it is the only framework which is flexible enough for their implementation [12, 13, 17, 18, 24, 29]. *This state of affairs forces stream processing application developers in a problematic position: they must choose between using the rigid high-level programming model offered by tools such as Spark or Flink (at the risk of the performance of their application being hampered by the hardwired distribution strategy imposed by their DSPF of choice), or they use the programming model offered by Storm to gain control over the distribution of their application (at the cost of having to deal with the substantial complexity imposed by Storm's programming model).*

In this paper, we present a novel programming model, called *Skitter* (Section 4). Skitter is a *dual* model: it introduces separate programming abstractions for the expression of data processing logic and distribution strategies. Separating the definition of these concerns allows both of these concerns to be modified independently. The model to express distribution strategies is *open*: expert developers can use the provided abstractions to write and adapt distribution strategies, allowing novel or existing strategies to be implemented as needed. Thus, Skitter offers a rigid, high-level programming model which still enables developers to select the most performant distribution strategy for each operation in their application. The set of abstractions we introduce through Skitter, which enable the implementation of distribution strategies in a modular fashion, form the main contribution of this paper.

Skitter is a domain-specific language built on top of Elixir[1] (Section 5). Validating our work is done by implementing several existing distribution strategies [8, 13, 17, 18] from the literature in both Skitter and Storm, after which we reproduce the benchmarks described in these papers. These benchmarks serve as the basis for both a quantitative and a qualitative evaluation. Our qualitative evaluation discusses the implementations of both benchmarks and argues that Skitter enables the implementation of distribution strategies from the state of the art in a modular fashion. Our quantitative evaluation compares the performance of both implementations and shows that distribution strategies expressed in Skitter exhibit the expected performance characteristics and that the throughput rate of these applications falls within the same order of magnitude as those obtained by Storm. Our performance evaluation also compares the performance of the Skitter benchmarks with an ad-hoc implementation of the same benchmarks to measure the overhead introduced by Skitter's abstractions. Finally, we compare the behavior of several distribution strategies and show that the most performant strategy for a given operation is situational. Thus, we show that the performance of a stream processing application can be improved by selecting the most appropriate distribution strategy and that Skitter enables this without polluting the data processing logic of the application with accidental complexity.

## 2 Background

Before we present our solution, we briefly discuss DSPFs and how they distribute stream processing applications over a cluster. We do this based on an example, namely a stream processing application calculating the conversion rate of ads, which we use throughout the paper. This cluster application receives a stream of clicks, representing users clicking on ads, and a stream of sales, representing users purchasing products. Based on these inputs, the application calculates how many of the clicks on a given ad result in a sale; this value is recalculated as new data enters the application and published to some external system (e.g. a dashboard).

**Programming model**    Stream processing applications are expressed by combining various data *sources* and data processing *operations*. Sources emit data on streams while operations ingest data from one or more streams, possibly emitting data on another stream. These connected sources and operations typically form a *Directed Acyclic Graph* (DAG). The DAG of our running example is shown in Figure 1. This application is implemented in terms of two sources: clicks and sales, and three operations: join, rate and publish. Both sources are linked to some external system and emit data records corresponding with sales or clicks into the rest of the application. The join operation ingests data from both sources: it buffers the clicks and sales that occur; when a sale of a particular product occurs within the same browser session where an ad for that product was clicked, it emits a data record indicating the ad that caused a sale to occur. The rate operation ingests data emitted by the join operation and the clicks source; it maintains two counters for each ad: one counting the number of times the ad has been clicked and another counting the number of times clicks on the ad resulted in a sale. Each time the rate operation receives a data record for an ad, it updates the counters associated with the ad, after which it emits an updated conversion rate for the ad. The conversion rate is received by the publish operation, which sends it to an external system.

---

[1] https://elixir-lang.org/, visited on 2025-01-21.

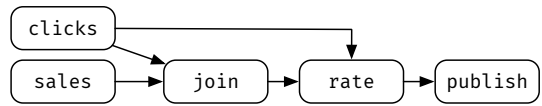

**Figure 1** DAG of a stream processing application calculating ad conversion rates.

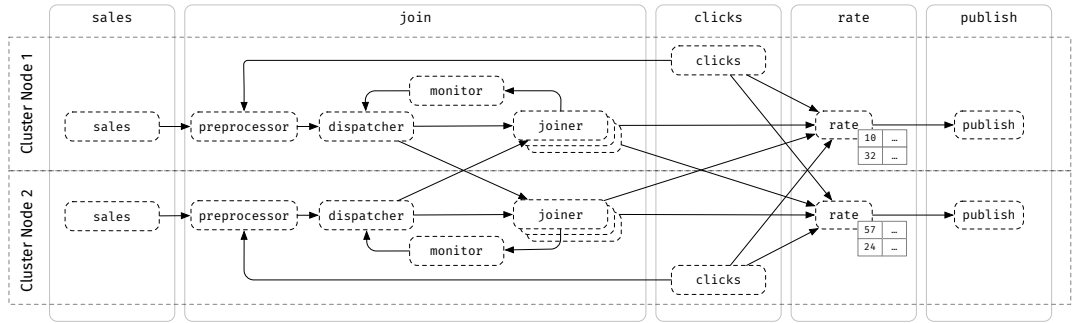

**Figure 2** Possible run-time view of the application shown in Figure 1 distributed over two cluster nodes. Each row depicts a cluster node; columns depict the sources and operations of the application.

**Distribution strategies**   To run a stream processing application on a cluster, its operations and sources must be distributed over the cluster nodes. This is done by the distribution strategy of the operation or source which makes the following decisions:
- How many *workers* are spawned to process data for the operation or source?
- Which information is exchanged among these workers?
- What is the exact task of each of these workers?
- How is the state of the operation or source partitioned among these workers?

Figure 2 visualizes a *potential* deployment of our ad conversion rate application distributed over a cluster with two nodes (for the sake of the example). Below, we discuss the strategies used to distribute the operations and sources that constitute the application.

**publish** The strategy for publish spawns a worker on every cluster node, as shown in Figure 2. Since the publish operation is stateless, no state is partitioned among these workers. When the publish operation needs to process a data record, it is sent to the worker of this operation on the current cluster node, which processes the data record, sending it to an external system.

**clicks and sales** These sources are stateless and distributed similarly to publish.

**rate** Similarly to publish, a worker is spawned on every cluster node to process data for the rate operation. Unlike publish, the rate operation is stateful, as it needs to maintain two counters for every ad. These counter pairs are partitioned over both of the operation's workers: when an ad click or sale needs to be processed, the distribution strategy needs to ensure that the event is correctly dispatched to the worker which maintains the counters for this particular ad. The worker which receives the event updates the relevant counter for the ad, after which it calculates the new conversion rate for the ad and emits it.

**join** The join operation uses the so-called "FastJoin" strategy [29]. We do not explain this strategy here for the sake of brevity. We do draw attention to the fact that this distribution strategy spawns different worker types (i.e. preprocessor, dispatcher, monitor and joiner), each with their own specific purpose. While other strategies exist for the join operation (e.g. [8, 13, 28]), we specifically include FastJoin here to showcase a complex distribution strategy which requires multiple worker types.

Distribution strategies determine *how* the work to be performed is balanced over the nodes of a cluster and which communication occurs between these nodes. A balanced distribution of work is essential to utilize the parallelism of a cluster environment, while sending a message over the network between cluster nodes introduces a significant amount of overhead. As such, distribution strategies are key to the performance of a distributed stream processing application.

Unfortunately, there is no one-size-fits-all distribution strategy for a given operation. This is the case as the performance of a strategy is impacted by the properties of its operation [18], the properties of the data it receives [9, 28] and the properties (e.g. heterogeneity) of the cluster on which the application is executed [18, 26]. In spite of this, modern DSPFs make it difficult to select an appropriate distribution strategy for an operation in a stream processing application, as we discuss in the next section.

## 3 Distributed Stream Processing Frameworks

Modern DSPFs offer a programming model in which stream processing applications are built by combining well-known operators such as map, reduce and filter. Listing 1 depicts an implementation of the running example in such a model. We discuss how the operations shown in Figure 1 are represented in this implementation below:

**clicks and sales** Sources which receive events corresponding with ad clicks and sales are created on lines 1 and 2, respectively.

**join** Lines 4 to 7 implement the join operation which matches events from both sources with one another to find ad clicks that resulted in a sale. Sales of a product are paired with clicks on ads related to that product if the click occurred in the same browser session as the sale. The **where** and **equalTo** operators are used to extract a key from the sales and clicks streams, respectively. In both cases, the product_id and browser session are extracted from the received event to form a tuple which serves as the key. The **apply** operator receives an ad click matched with a sale and produces a {:sale, ad_id} tuple, indicating the ad resulted in a sale.

**rate** DSPFs typically feature a **reduce** operator to perform any aggregations. However, such an operator only accepts a single input. Therefore, on line 8, we use the **union** operator to combine the stream resulting from the join operation with the stream of ad events. We use the **map** operator to wrap these events in a {:click, ad_id} tuple in order to distinguish ad clicks and sales from one another. The KeyBy operator on line 9 is used to group the various incoming data records for **reduce** by key. In this case, we group the events based on their ad_id. On lines 10 to 12, the **reduce** operator is used to update the amount of sales or clicks for an ad, depending on the received event. After each reduction, the current state for the ad is emitted. On line 14, the **map** operator is used to calculate the current conversion rate of the ad, based on the numbers of clicks and sales.

**publish** Each conversion rate emitted by the **map** operator is published on line 15.

Developers using such a high-level model only need to specify the data processing logic of their application as the DSPF distributes each operator over the cluster automatically. This implies that the developer retains very little control of the distribution of each operator, as all distribution logic is handled by the DSPF. To alleviate this issue somewhat, most high-level DSPFs offer built-in distribution operators which can be used to configure the behavior of the distribution strategies of the DSPF. For instance, both Spark and Flink introduce operators which can be used to specify the amount of parallel instances spawned for an operation; Flink's join operator also takes an optional argument which configures the behavior of its built-in distribution strategy.

However, these operators do not offer developers control over how the computations of an operator are distributed over the cluster or over the communication between these computations. In other words, these operators enable developers to slightly configure existing distribution strategies, but do not allow them to select a radically different strategy to use for an operator or to create new strategies.

As a consequence, developers who aim to implement a distribution strategy often use Apache Storm [23] for this purpose [12, 13, 17, 18, 24, 29], as it is the only framework flexible enough for their implementation. Storm applications are written by building a graph, called a *topology* in Storm, which consists of various sources, called *spouts*, and operations, called *bolts*. Spouts and bolts can be connected to each other through the use of *groupings*, which represent edges. An example of a Storm topology can be seen in Listing 2, which depicts a Storm topology implementing the running example. This code is essentially a textual description of the *run-time* DAG shown in Figure 2: each worker type shown in Figure 2 is added to the application's topology through the use of **setBolt** or **setSpout**. Bolts and spouts are instantiated with

**Listing 1** Pseudocode representing a high-level DSPF implementation of the running example.

```
1  clicks = source()
2  sales = source()
3
4  sales.join(clicks)
5      .where(fn sale -> {sale.session, sale.product_id} end)
6      .equalTo(fn ad_click -> {ad_click.session, ad_click.product_id} end)
7      .apply(fn ad_click, sale -> {:sale, ad_click.ad_id} end)
8      .union(clicks.map(fn ad_click -> {:click, ad_click.ad_id} end))
9      .keyBy(fn {_, ad_id} -> ad_id end)
10     .reduce({0,0}, fn
11         {:click, ad_id}, {clicks, sales} -> {ad_id, {clicks + 1, sales}}
12         {:sale, ad_id}, {clicks, sales} -> {ad_id, {clicks, sales + 1}}
13      end)
14     .map(fn {ad_id, {clicks, sales}} -> {ad_id, sales / clicks} end)
15     .publish()
```

**Listing 2** Simplified version of a Storm topology implementing the running example.

```
1  builder = new TopologyBuilder();
2
3  builder.setSpout("sales", new SalesSpout(), 2)
4  builder.setSpout("clicks", new ClicksSpout(), 2)
5
6  builder.setBolt("join-preprocessor", new PreProcessBolt(), 2)
7         .localGrouping("clicks")
8         .localGrouping("sales")
9  builder.setBolt("join-dispatcher", new DispatcherBolt(), 2)
10        .globalGrouping("join-preprocessor")
11        .globalGrouping("join-monitor")
12 builder.setBolt("join-joiner", new JoinBolt(), 8)
13        .customGrouping("join-dispatcher", new FastJoinGrouping())
14
15 builder.setBolt("rate", new RateBolt(), 2)
16        .fieldsGrouping("clicks", "ad-id")
17        .fieldsGrouping("joiner", "ad-id")
18 builder.setBolt("publish", new PublishBolt(), 2)
19        .localGrouping("rate")
```

an object which encapsulates the logic performed by the bolt or spout when it receives data, and with a *parallelism hint*, which determines how many parallel copies of these bolts and spouts are spawned at run-time.

Storm *groupings* are responsible for linking spouts and bolts to each other and also determine which parallel copy of the receiving bolt will process the received data at run-time. Storm defines several built-in groupings (such as the **localGrouping** shown on lines 7 and 8), but also offers a generic interface, **customGrouping**, for the definition of custom groupings, as shown on line 13. In this way, Storm offers developers full control over where each data record processed by their application is sent and processed.

**Skitter: A DSPF with Pluggable Distribution Strategies**

The programming model offered by Storm allows for the implementation of distribution strategies. However, it is significantly more complex than the one introduced by high-level DSPFs and was not designed to express distribution strategies in a modular fashion. Concretely, three key hurdles make it difficult to express and modify distribution strategies in Storm.

**Scattered distribution logic** A distribution strategy must specify how many workers are created for an operation, how the state of this operation is partitioned over these workers, and how information is exchanged among them. In Storm, the number of workers created is specified in the topology definition, while state is handled in spout and bolt classes; communication between workers is handled in grouping definitions. Thus, the definition of a distribution strategy is scattered over several locations, making it harder to modify. In Listing 2, this is most obvious in the definition of the FastJoin strategy (lines 6–13), the definition of which is scattered across three different bolt classes, one grouping class and the application's topology.

**Tangled distribution and application logic** Storm does not distinguish application from distribution logic. In simple cases, groupings handle the distribution logic of an operation, while bolts and spouts handle the application logic. In practice, however, complex distribution strategies often require distribution-level logic to be present in bolts or spouts. As a result, distribution and application-level logic is often entangled in Storm, making it hard to modify either [5]. This is also clearly visible in the definition of the FastJoin strategy in Listing 2: the PreProcessBolt, DispatcherBolt, FastJoinGrouping and the various grouping declarations (i.e. uses of **localGrouping** and **globalGrouping**) define distribution logic while the JoinBolt defines both distribution and application logic.

**Lack of support for creating different task types** Complex distribution strategies often require the creation of different workers, each of which performs a different task for the distribution strategy. In Storm, any task spawned for a spout or bolt executes identical logic when invoked. As a result, distribution strategies that require different worker types can only be expressed in Storm by introducing multiple spouts, bolts and groupings, thus polluting the definition of the application's topology and worsening both the scattering of distribution logic and its entanglement with application logic. This issue is also visible in Listing 2, which describes the runtime DAG of the application (shown in Figure 2), rather than its application DAG (Figure 1).

**Conclusion**

- High-level DSPFs provide an operator-based programming model which makes it easy for developers to specify the data processing logic of their applications. However, each of these operators is strongly tied to its distribution strategy, which makes it impossible for developers to pick another distribution strategy.
- Storm offers a low-level model which enables developers to implement their own distribution strategies, but is not designed for this purpose, which leads to several issues that make it hard to implement and modify distribution strategies in Storm.

▪ **Listing 3** Workflow definition of the running example.

```
1  workflow do
2    node(ClicksSource, as: clicks)
3    clicks.out ~> join.right
4    clicks.out ~> rate.clicks
5
6    node(SalesSource, as: sales)
7    ~> node(Join, with: FastJoin, as: join)
8    ~> node(Rate, with: KeyedState, as: rate)
9    ~> node(Publish)
10 end
```

In the next section, we introduce Skitter, which aims to combine the ease-of-use of high-level DSPFs with the expressive power and flexibility offered by Storm.

## 4  The Skitter Programming Model

In Skitter, stream processing applications are implemented using three main abstractions: *workflows*, *operations* and *strategies*. Operations define the data processing logic of an application, while strategies define the distribution logic for an operation. Workflows combine operations and their associated strategies into a DAG to create a stream processing application. We discuss each of these concepts below.

### 4.1  Workflows

A Skitter workflow is defined as a DAG consisting of several connected *nodes*. Each of these nodes represents a single data processing step and embeds an operation (i.e. the data processing logic of the step) and the strategy associated with this operation (i.e. the distribution logic of the data processing step); optionally, arguments can be passed to a node. Nodes are connected through *links*, which wire the in- and outputs of the various nodes to each other. In this way, stream processing applications can be built by combining several operations into a DAG.

Listing 3 defines our running example as a Skitter workflow. This workflow definition is a textual representation of the application DAG shown in Figure 1. The various sources and operations which define the data processing steps of the application are defined as nodes, while data dependencies between the nodes are expressed through the use of the link operator, `~>`. The link operator can be used to link a specific output of a node to the input of another node, as on lines 3 and 4, or it can be used to "chain" nodes after each other, as shown on lines 7 to 9; when this is done, the first output of a node is linked to the first input of the next chained node.

**Specifying the distribution strategy of a node**   The main goal of Skitter is to enable developers to select distribution strategies for their operations. Therefore, the **with:** operator can be used to specify which strategy is used to distribute the operation of

**Skitter: A DSPF with Pluggable Distribution Strategies**

◼ **Listing 4** Operator-style definition of a word count application.

```
1  workflow do
2    source()
3    ~> flatmap(fn text -> String.split(text) end, with: RepartitionedOutput)
4    ~> keyed_reduce(fn word -> word end, fn count -> count + 1 end, 0)
5    ~> print()
6  end
```

a particular node. For instance, on line 7, we specify that the `FastJoin` distribution strategy has been chosen to distribute the `Join` operation. In many cases, developers may not want to tweak the distribution strategy of an operation. Therefore, operations may feature a default distribution strategy which is used when **with:** is omitted.

**Well-known generic operators** High-level DSPFs offer various well-known operators which can be used to build stream processing applications. While convenient, these operators can be restrictive when an application does not fit into their constraints. For instance, in Listing 1, we used the **union** and **map** operators to merge two steams because the **reduce** operator can only operate on a single stream. To avoid this issue, nodes in a Skitter workflow are created with an operation, similar to how spouts and bolts are instantiated with an object in Storm. However, Skitter provides several operations, such as `Map`, `Filter` and `KeyedReduce` which behave like these well-known operators. Moreover, Skitter introduces syntactic sugar for defining nodes which use these operations; these can be mixed with regular node definitions which makes it possible to write (parts of) a stream-processing application in the operator style. An example of this can be seen in Listing 4, which uses the operator style to define a workflow which counts words.

### 4.2 Operations

A Skitter operation defines a conceptual data processing step which can be embedded in a workflow. At run-time, an operation is distributed by some distribution strategy. Operation definitions consist of meta-information, which is used to embed the operation inside a workflow, and of various *callbacks*, which implement the data processing logic provided by the operation. Listing 5 shows the definition of the `Rate` operation used in Listing 3. Recall that this operation calculates the conversion rate of ads. It does this by maintaining two counters for each ad: one counting the number of times the ad has been clicked, and another counting the number of times clicks on the ad resulted in a sale. When the operation receives a click or sale event, it increments the appropriate counter after which it calculates and publishes the updated conversion rate for the ad.[2]

---

[2] Note that this operation can be implemented using operators, similar to how this was done in Listing 1. We provide an operation definition here instead for the sake of the example.

■ **Listing 5**  Definition of the Rate operation.

```
1  defoperation Rate, in: [sales, clicks], out: conversion_rate, strategy: KeyedState do
2    initial_state {0, 0}
3
4    defcb key(data), do: data.ad_id
5
6    defcb react(data) do
7      {clicks, sales} = state()
8      {new_clicks, new_sales} = case port_of(data) do
9        :sales -> {clicks, sales + 1}
10       :clicks -> {clicks + 1, sales}
11     end
12     state <~ {new_clicks, new_sales}
13     {data.ad_id, new_sales / new_clicks} ~> conversion_rate
14   end
15 end
```

The first line of the operation definition defines the meta-information of the operation: it specifies that this operation responds to two input streams: sales and clicks, and that it produces a single output stream, conversion_rate. It also specifies the default distribution strategy of the operation, KeyedState. The second line in Listing 5 specifies the initial state of the operation, which can be retrieved by the strategy when needed, as we discuss later. The initial state of the Rate operation is a tuple containing two counters, each starting at 0. Lines 4 and 6–14 define the callbacks of the Rate operation; the key callback obtains a key for an incoming data element while the react callback defines how a data element is processed. We discuss both callbacks in depth in the following paragraphs.

**Required callbacks**  The data processing logic of an operation is implemented in terms of callbacks. In the most simple case, an operation only needs to define a single callback which specifies how the operation responds to incoming data records (such as the react callback). However, a distribution strategy often requires several strategy-specific callbacks to be present. For instance, the KeyedState distribution strategy used by Rate needs to ensure each ad is processed by the worker which maintains the counters for this particular ad, as discussed in Section 2. Typically, a key is used to identify this worker. However, obtaining this key for a particular data record belongs to the data processing logic of an application. Therefore, the KeyedState distribution strategy requires an operation to implement a key callback, which extracts such a key from an incoming data record. Embedding this logic inside the operation definition ensures that the KeyedState strategy can remain agnostic to the data processing logic of the operation it distributes. The key callback, defined on line 4, obtains the key of an incoming data record by extracting its ad_id.

**Callback definition**  The react callback, defined on lines 6 to 14, specifies the data processing logic that is invoked when the Rate operation receives a new data record. This operation needs to update the state associated with a particular ad, and publish the new conversion rate obtained from the updated state to downstream operations. Managing state, however, is a concern to be managed by distribution strategies, as it determines where computations of an operation may be executed. For instance, we have already discussed that the state of the Rate operation is partitioned over several workers. Similarly, a distribution strategy should have control over when an operation can send data to its downstream operations (e.g. to execute a computation multiple times, as in [24]). To reconcile these concerns, Skitter conceives callbacks as functions which accept a state along with their regular arguments and which return a potentially updated state, emitted data and a return value. In other words, a callback is a function, $\gamma : state, args \mapsto state', retval, emit$. Distribution strategies can then call these callbacks, providing the state on which the callback may operate.

The KeyedState strategy calls the react callback with the state associated with a particular ad. Inside the body of a callback, the **state**() operator may be used to access this state, as on line 7. On lines 8 to 11, the callback updates either the clicks or the sales counter, depending on whether a value was received on the sales or clicks input stream. The updated counters are then stored as the new state, through the use of the **<~** primitive. Finally, the operation calculates the new conversion rate of the ad, and publishes it along with the ad_id through the use of the *emit* operator, **~>**. The callback DSL ensures the emitted values, updated state and return value of the callback are returned to the strategy, which we discuss below.

### 4.3 Distribution Strategies

Skitter introduces distribution strategies as a means for expert (meta) programmers to specify how an operation is distributed over a cluster. A distribution strategy is implemented by defining a set of *hooks*, called by the runtime system in response to predefined events. Writing these hooks involves dealing with *workers*, which are computational entities based on the actor model [1], which maintain state and perform computations for operations. Distribution strategies are implemented by defining the *deploy*, *deliver* and *process* hooks:

**deploy**  Creates an initial deployment of the operation over the cluster, typically by spawning workers.

**deliver**  Delivers data emitted by an upstream operation to a worker of this operation.

**process**  Processes a single value received by a worker of this operation.

We discuss these hooks using our running example. We first provide a brief overview, after which we provide a detailed discussion in the paragraphs below. An overview of the hooks, the events that trigger them, and their purpose is provided in Table 1.

Listing 6 shows the definition of the KeyedState strategy, used to distribute the Rate operation of the running example. This strategy partitions the state of an operation based on a set of keys: inside the deploy hook, a worker is created for each cluster node; each of these workers maintains the state of several keys. When the operation receives a new data record, the deliver hook is invoked; inside this hook, the key of the received record is extracted and hashed. Based on this hash, the data record is relayed to the appropriate worker. When this worker receives the data record, the process

■ **Listing 6** Definition of the KeyedState distribution strategy.

```
defstrategy KeyedState do
  defhook deploy(args) do
    Remote.on_all_workers(fn -> local_worker(Map.new(), :aggregator) end)
    |> Enum.map(fn {remote, worker} -> worker end)
  end

  defhook deliver(data) do
    key = call(:key, args: [data]).result
    aggregators = deployment()
    idx = rem(Murmur.hash_x86_32(key), length(aggregators))
    worker = Enum.at(aggregators, idx)
    send(worker, data)
  end

  defhook process(data, state_map, :aggregator) do
    key = call(:key, args: [data]).result
    state = Map.get(state_map, key, initial_state())
    res = call(:react, state: state, args: [data])
    emit(res.emit)
    Map.put(state_map, key, res.state)
  end
end
```

hook is invoked, which will fetch the state associated with the received data record and call the react callback of the operation with this state. The state and emitted data returned by react will be used as the new state for the received key and sent to downstream operations, respectively. We discuss every hook and its implementation in the KeyedState strategy in detail in the following paragraphs.

**Initial deployment over a cluster**   The deploy hook is responsible for distributing an operation over the cluster nodes. When a Skitter application is started, the Skitter runtime system calls the deploy hook of the distribution strategy of each operation in the application with the arguments passed to the workflow node. The deploy hook is responsible for creating the resources required for the operation to respond to incoming data records. These resources include references to workers, which will perform work for the operation, and any constants maintained by the operation (e.g. a function provided to a higher-order operator such as map). The result of the deploy hook is returned to the runtime system which stores it as the so-called deployment data. This data is automatically copied to every node in the cluster and can be accessed by the deliver and process hooks. Typically, the deploy hook creates a set of workers, after which references to these workers are returned to be stored inside the deployment data.

In our example, the deploy hook is used to create one worker for each worker node on the cluster (line 3). This is done through the use of **on_all_workers**, which calls a provided function on every worker node in the cluster, and through the use of **local_worker**, which creates a worker with an initial state (an empty map), and a *role* (which we discuss later). On line 4, we transform the result of **on_all_workers** (a mapping of node names to worker references) into a list of worker references, which is returned to be stored in the deployment data.

**Delivering data to workers**  Skitter's workers perform computations and manage state for the distribution strategy. The deliver hook is responsible for selecting which worker a data record is sent to. When an operation's strategy emits data, the deliver hook of the strategy of each of its downstream operations is called. The hook must then make sure that the received data record is sent to a worker, where it can be processed. Since a node in a Skitter workflow may select any distribution strategy for an operation in their application, the deliver hook cannot make any assumption about the current location of the data to be sent. Typically, the deliver hook selects a worker from the deployment data to send the data to.

The deliver hook of our example calls the *key* callback of its operation to obtain the key associated with the incoming data record (line 8). Afterwards, it fetches the worker references from the deployment data through the use of the **deployment** primitive (line 9). Next, the key is hashed and reduced modulo the amount of workers to obtain an index (line 10), which is used to obtain a worker reference (line 11). Finally, the **send** primitive is used to send the data to the selected worker (line 12).

**Processing data**  The **send** primitive can be used to send a message to a worker. Messages sent to a worker are stored in a queue and are processed sequentially in the order of arrival. When a worker receives a message, the process hook is called with the message, the current state of the worker and the worker's role. This hook specifies how a worker processes a single message in its queue. The value returned by this hook is used as the new state of the worker. Typically, the process hook invokes a callback of the operation to process the received data with the appropriate state.

In our example, the process hook calls the *key* callback of the operation to obtain the key associated with the incoming data record (line 16). This key is then used to fetch the state of this particular key from the worker's state, a map. When the worker's state does not contain a value for the key, the initial state of the operation is used instead (line 17). Next, the *react* callback of the operation is called, with the received message and the obtained state, to process the incoming data record (line 18). The value emitted from the operation is emitted through the use of the **emit** primitive (line 19), which causes the Skitter runtime system to call the deliver hook of any downstream operations. Finally, the updated state returned from calling *react* is inserted into the state map and returned as the result of the process hook, which will store it as the state of the worker.

The deploy, deliver and process hooks discussed above are the key mechanism through which Skitter enables the expression of distribution strategies. The purpose of each of these hooks is summarized in Table 1, while the primitives available for use inside these hooks is shown in Table 2. In the following paragraphs, we discuss two additional features, which enable the implementation of complex distribution strategies in Skitter.

▪ **Table 1** Summary of the hooks that a distribution strategy needs to implement.

| Hook | Arguments | Return | Event | Goal |
|---|---|---|---|---|
| deploy | $args^1$ | $dep^2$ | Application start. | Distribute operation. |
| deliver | $data$ | | $data$ emitted upstream. | Send $data$ to worker. |
| process | $msg, state, role$ | $state'$ | Worker receives $msg$ | Process $msg$, update worker's $state$. |

[1] Arguments passed to the workflow node.
[2] The deployment data for the current operation.

In Section 2, we describe how the FastJoin distribution strategy creates several distinct worker types. Later, we discussed that such a distribution strategy can only be expressed in Storm by defining several bolts and groupings, scattering the logic of the strategy over several classes and polluting the application's topology by wiring these strategy-specific bolts and groupings to each other. Skitter supports the creation of several different worker types within a single strategy to avoid scattering distribution logic over several locations and explicitly differentiates between inter- and intra-strategy communication to ensure no strategy-specific data dependencies are required in a workflow definition.

**Different worker types** Skitter workers are created with a *role*, which is passed as an argument to the process hook. This role can be used to execute different logic depending on the purpose of the particular worker. In our implementation, pattern matching can be used to define several role-specific clauses of the process hook.

**Inter-strategy communication** Skitter distinguishes between intra-strategy communication and inter-strategy communication. The former is handled through the use of **send**, which requires a reference to a worker, which is strategy-specific, while the latter is handled through the use of **emit**, which causes the runtime system to call the appropriate deliver hooks. This distinction ensures that strategies can remain agnostic to each other, ensuring an appropriate distribution strategy can be selected for each node in the application.

## 5 Implementation

We implemented Skitter as a domain-specific language built on top of Elixir. The DSL is open-source, well-documented and available online.[3] Elixir was chosen over more traditional platforms such as the JVM due to its excellent built-in support for meta-programming and support for cluster-oriented distribution inherited from Erlang.

---

[3] https://github.com/mathsaey/skitter, visited on 2025-01-21.

- **Table 2** Core primitives available inside hooks.

| Primitive | Arguments | Description |
| --- | --- | --- |
| **call** | $cb, state, args$ | Call callback *cb* with *state* and *args*. |
| **initial_state** | | Fetch the initial state of the operation. |
| **deployment** | | Fetch the deployment data, cannot be used in the deploy hook. |
| **local_worker** | $state_0, role$ | Create a worker on the local node with initial state $state_0$ and the given *role*. |
| **remote_worker** | $state_0, role$ | Create a worker on a remote node with initial state $state_0$ and the given *role*. |
| **send** | $ref, msg$ | Send *msg* to the worker with reference *ref*. |
| **emit** | $data$ | Publish *data* to downstream operations. |
| **self** | | Get a reference to the current worker. |
| Remote.**on** | $remote, f$ | Execute function $f$ on *remote*, a worker node. |
| Remote.**on_n** | $n, f$ | Execute function $f$ on $n$ worker nodes. |
| Remote.**on_all_workers** | $f$ | Execute function $f$ on all worker nodes. |
| Remote.**self** | | Get the name of the current cluster node. |

Skitter defines macros to be used to define operations, strategies and workflows. These macros are used to transform a programmer's operation and strategy definitions into Elixir module definitions; workflows are transformed into a data structure representing the DAG of the stream processing application.

The implementation also provides a runtime system which enables the distributed execution of Skitter applications on a cluster. The Skitter runtime system calls the right hooks to deploy nodes or when data is emitted or received by workers. It also defines workers (which are implemented on top of Elixir actors) and the primitives shown in Table 2 (which enable developers to interact with the runtime system).

Apart from the DSL and its supporting runtime system, Skitter also provides several predefined operations and strategies, along with syntactic sugar which enables stream processing applications to be written in the operator style. Due to the aforementioned meta-level interface provided by Skitter, these built-in operations and their strategies required no changes to Skitter's implementation. Instead, these operations and their strategies were defined using the abstractions described in the previous section.[4]

# 6 Evaluation

In this section, we evaluate Skitter by implementing several distribution strategies published in the Big Data stream processing community and by reproducing their benchmarks in Skitter. Based on these benchmarks, we aim to answer the following questions:

---

[4] Developers can also extend Skitter with their own — custom — operators.

▬ **Table 3** Operations, distribution strategies and configurations used in the evaluation.

| Operation | Strategy | Configuration | Label |
| --- | --- | --- | --- |
| WordCount | D-Choices [18] | 80 workers | D-C |
|  | W-Choices [18] | 80 workers | W-C |
|  | Partial Key Grouping [17] | 80 workers | PKG |
|  | Key Grouping | 80 workers | KG |
|  | Shuffle Grouping | 80 workers | SG |
| Join | Join-Matrix [8] | 20 workers in a 4 × 5 matrix | JM |
|  | Join-Biclique [13] | 20 workers | JB |
|  | Join-Biclique ContRand [13] | 20 workers, 5 subgroups | JB-CR |

**Q1** Does Skitter enable the expression of distribution strategies in a modular fashion?
**Q2** Does Skitter influence the performance characteristics of distribution strategies?
**Q3** Do the Skitter language abstractions introduce a significant amount of overhead?
**Q4** Can application performance be improved by selecting an alternative strategy?

## 6.1 Experimental Setup

We answer our research question based on an implementation of the benchmarks described by Lin, Ooi, Wang, and Yu [13] and Nasir, Morales, Kourtellis, and Serafini [18]. The former work introduces distribution strategies for join operations, while the latter introduces distribution strategies for key-based reduce operations. We selected these particular works for several reasons:

- join and reduce operations are commonly used in stream processing applications and can have a major impact on their performance [11].
- These works compare several different distribution strategies for the same operation with each other in a single experiment.
- The benchmarks described in these works operate on data that is freely available or that can be generated synthetically.

Before describing our experiments, we elaborate on the benchmarks and the distribution strategies compared therein.

**Benchmarks**  The benchmarks compare the throughput of various distribution strategies distributing the same operation. Table 3 provides a summary of the operations, strategies and their configurations.

**WordCount**  Nasir, Morales, Kourtellis, and Serafini [18] introduce distribution strategies for key-based stateful operations geared towards handling highly-skewed workloads (i.e. workloads where the majority of the incoming data records are associated with a small set of keys). This work introduces the *D-Choices* (D-C) and *W-Choices* (W-C) strategies for key-based reduce operations [18] and compare them to the *Partial Key Grouping* (PKG) strategy, introduced by the same authors in earlier work [17], and to the *Key Grouping* (KG) and *Shuffle Grouping* (SG) strategies.

The Key Grouping strategy refers to the use of hash-based grouping, similar to the `KeyedState` strategy described in Section 4.3. Partial Key Grouping splits the processing of data elements for each given key over two workers to remain resilient to skewed workloads. W-Choices and D-Choices take this idea further by distributing the processing of the most frequent keys over several (D-C) or all (W-C) workers. The Shuffle Grouping strategy randomly distributes the processing of data elements over the cluster and is used as a comparison point, as it achieves a uniform distribution of work over the cluster, regardless of skew.

The authors investigate if their strategies attain an even distribution of work over the cluster. This is done by evaluating these strategies in the context of an application which counts random words drawn from a synthetically generated highly skewed dataset. The dataset contains $10^4$ words, distributed according to a Zipf distribution with an exponent $z \in \{1.4, 1.7, 2.0\}$; higher values for $z$ indicate a higher level of skew. The application does not perform any operations on the aggregated state. Instead, work is simulated by adding a fixed delay (of 1ms) to the processing of each message.

The W-Choices, D-Choices and PKG strategies all split the processing of data for a single key over several cluster nodes. However, since the authors are only interested in the even distribution of work over the cluster, the split partial results are never merged again, as they would be in a real application.

**Join** Lin, Ooi, Wang, and Yu [13] introduces the *Join-Biclique* (JB) strategy and its *Join-Biclique ContRand* (JB-CR) variant for the distribution of the join operation and compares these to the *Join-Matrix* (JM) strategy [8].

The Join-Matrix strategy stores each element to be joined several times, which can prove troublesome when joining large amounts of data. The Join-Biclique and Join-Biclique ContRand strategy only store each data element once, but require extra synchronization logic to ensure correctness. The Join-Biclique strategy and its ContRand variant differ in how they distribute data over the available join workers: the Join-Biclique strategy distributes data over workers randomly, requiring potential matches to be processed in more places while the ContRand variation uses a key-based approach to avoid this at the trade-off of becoming sensitive to skewed input data.

Lin, Ooi, Wang, and Yu [13] compare these strategies by running two join queries present in the TPC-H benchmark suite.[5] Specifically, the join operations present in queries five and seven were used. For query 5, this is (Region ⋈ Nation ⋈ Supplier) ⋈ LineItems, while (Nation ⋈ Supplier) ⋈ LineItems is used for query 7. The input for this experiment is generated using the `dbgen` tool, which is part of the benchmark suite. Before performing the experiments, the data is processed to only include the values used by the application. Moreover, input data is fed to the application at a fixed rate, which differs for each input stream.

---

[5] https://www.tpc.org/tpch/, visited on 2025-01-21.

We implement all the distribution strategies and their benchmarks mentioned above in both Skitter and Storm, and run the experiments described in the original work. The implementations of these strategies and the accompanying benchmark code are available as an artifact submitted along with this paper [20]. For both experiments, we adjust the amount of created workers (or tasks for Storm) to fit the cluster environment on which we evaluate our work (discussed below). For the WordCount experiment, we create one worker for each hardware thread of the worker nodes of the cluster, leading to a total of 80 workers. For the Join experiments, we aim to spawn an equal amount of workers on every cluster node, while still obtaining a reasonable distribution of work for the Join-Matrix strategy. For this reason, we spawn 20 workers for each join operation in the Join experiments. Table 3 provides a summary of the strategies and configurations considered in our evaluation.

In our performance experiments, we consider the average throughput of the applications. The average throughput is calculated by measuring the elapsed time between the first and final data element and dividing this time by the total amount of processed data elements. For the join experiments, input is provided to the application at a fixed rate. A limitation of this setup is that the input rate may not be sufficient to saturate the application, in which case the measured throughput rate may be lower than the maximum throughput rate of the application. We work around this limitation by, for every experiment, experimentally determining the highest input rate for which the experiment consistently finishes within ten minutes (for the 10 GB experiments) or one hour (for the 80 GB experiments) after receiving all inputs.

In Storm, programmers typically programmatically acknowledge emitted data records which enables the framework to ensure each record is processed at least once. The implementation of Skitter does not provide such a mechanism. Therefore, we do not use this mechanism in the Storm implementation of the experiments for the sake of a fair comparison.

**Environment**   All of our experiments are performed on a cluster of ten worker nodes and one master node. Each node is equipped with a 4-core Intel® E5–1620 Xeon® CPU with 8 hardware threads running at 3.50 GHz. Every machine has 32 GB of 2133 MHz DDR4 RAM, and is configured with 32 GB of swap space. The nodes are connected with a 10 Gigabit Ethernet connection and run Ubuntu 22.04.4. We compiled and ran our evaluation with Elixir 1.17.2 on Erlang/OTP 27.0.1 and with Storm 2.6.3 on OpenJDK 21.0.3.

### 6.2  Implementation in Skitter

Before discussing the results of our evaluation we briefly discuss how Skitter impacted the implementation of the various distribution strategies. We refer readers interested in the full details of the implementation to our evaluation artifact [20] and the original work describing the strategies [8, 13, 17, 18].

**Stateful deliver logic**   The PKG, D-C and W-C distribution strategies all require stateful logic to decide which downstream worker will process a data record. In Storm, groupings (discussed in Section 3) can read and modify a state which is stored inside the task of the upstream operation. This is not the case in Skitter, where all state is managed by workers and where the deliver hook is stateless. Instead, the Skitter implementation of these strategies uses two different worker types: *forwarders* and *buckets*. The deliver hook sends a data record to be processed to a forwarder worker on the local node, after which its process hook can access the forwarder's state to execute the stateful logic required to select a bucket worker, which can process the data record.

**Complex worker interaction** The Join-Biclique strategy and its ContRand variant require complicated worker interaction to ensure the various workers responsible for joining data records do not emit duplicate values. Concretely, a set of *sender* workers needs to maintain a logical clock which needs to be synchronized and sent to all join workers at a regular interval. Again, we tackle this issue by introducing different worker types. Our implementation of the Join-Biclique strategy uses two worker types: *senders* and *joiners*. The first is responsible for sending data to joiners and for maintaining a logical clock which is sent to the joiners and other senders at regular intervals. The second is responsible for joining the received data records, when this is permitted by Join-Biclique's synchronization logic.

## 6.3 Results

### 6.3.1 Modularity of Distribution Strategies in Skitter

To examine the modularity of distribution strategies in Skitter, we measure which types of code a programmer must adjust to change the distribution strategy of an existing application and compare it to the types of code that need to be changed in the current state of the art, i.e. to Storm. This is done by implementing the experiments using a basic distribution strategy (KG for the wordcount experiment, JM for the join experiment), after which we modify the code to use a different distribution strategy and count the lines of code that were added or modified between both versions. Since we only changed the distribution strategy, the locations that were changed indicate which locations in the application a programmer must adjust to change distribution strategies. We categorize the changed lines according to the abstractions offered by the DSPF (Topology, Component and Grouping in Storm; Workflow, Operation and Strategy in Skitter), and the "Other" category, used to capture code expressed in the base language. The results of this experiment are shown in Table 4, where each row represents a strategy and every column represents the lines of code changed for each category in both Storm and Skitter compared to the implementation of the basic strategy. The results shown in this table are discussed per benchmark below.

**WordCount** The first two lines of Table 4 show that very little changes are needed to use the PKG or SG strategies in Storm. This is the case because Storm offers a built-in grouping for both of these strategies. Thus, this strategy can be used in Storm by making minor changes to the topology. Skitter does not include an implementation of these strategies by default, so changes are needed in two places: the strategy needs to

■ **Table 4** Lines of code added or modified to change distribution strategies compared to the KG (WordCount) and Join-Matrix (Join) implementation.

|  |  | Strategy | Storm | | | | | Skitter | | | | |
|---|---|---|---|---|---|---|---|---|---|---|---|---|
|  |  |  | Topology | Component | Grouping | Other | Total | Workflow | Operation | Strategy | Other | Total |
| WordCount | | SG | 1 | 0 | 0 | 0 | 1 | 1 | 0 | 8 | 0 | 9 |
| | | PKG | 1 | 0 | 0 | 0 | 1 | 1 | 0 | 46 | 0 | 47 |
| | | W-C | 1 | 0 | 29 | 91 | 121 | 1 | 0 | 71 | 25 | 97 |
| | | D-C | 1 | 0 | 59 | 91 | 151 | 1 | 0 | 107 | 25 | 133 |
| | | PKG† | 4 | 38 | 0 | 0 | 42 | 1 | 4 | 65 | 0 | 70 |
| | | W-C† | 4 | 38 | 29 | 91 | 162 | 1 | 4 | 90 | 25 | 120 |
| | | D-C† | 4 | 38 | 59 | 91 | 192 | 1 | 4 | 126 | 25 | 156 |
| Join | Q5 | JB | 29 | 162 | 46 | 0 | 237 | 3 | 0 | 119 | 0 | 122 |
| | | JB-CR | 29 | 162 | 61 | 0 | 252 | 3 | 0 | 134 | 0 | 137 |
| | Q7 | JB | 22 | 162 | 46 | 0 | 230 | 2 | 0 | 119 | 0 | 121 |
| | | JB-CR | 22 | 162 | 61 | 0 | 245 | 2 | 0 | 134 | 0 | 136 |

† With key merge logic.

be defined, and the workflow needs to be adjusted to use the new strategy (through the use of **with:**, described in Section 4.1).

The W-C and D-C strategies are not provided by Storm but can be defined as a grouping; the Skitter implementation of these strategies are expressed as a strategy used by the workflow, similar to how this was done for the PKG strategy. Table 4 shows that code in the "Other" category was added to implement these strategies in both Storm and Skitter. This code contains an implementation of the "Space-saving" algorithm [4, 15], used by the W-C and D-C strategies.

The WordCount benchmark is only concerned with how the various strategies balance work over the cluster. Therefore, it does not merge the partial results for each key generated by the PKG, D-C and W-C strategies. This merge logic would be present in a realistic application, however. The PKG, W-C and D-C rows marked with † compare an implementation of these strategies with this merge logic included compared to the base KG implementation (which does not split keys and therefore does not need this logic). In Storm, this results in the creation of an additional Component (a `MergeBolt`) and extra changes to the topology, as the merge bolt and the connections to this bolt need to be encoded in the topology definition. In Skitter, these changes are captured in the strategy definition, so no additional changes to the workflow are needed. Merging partial states associated with a key is application-specific behavior, however, which is therefore encoded as a callback inside of the word count operation. The changes to the operation are reflected in the PKG†, W-C† and D-C† rows of Table 4. Storing this logic inside the operation allows the reuse of the W-C, D-C and PKG strategies which stands in contrast to the merge bolt in Storm, which contains both distribution logic and application-specific behavior. The updated word count operation in the Skitter implementation is also still usable by the KG strategy, which can simply not call the added callback.

**Join** The Join benchmark consists of two queries: query 5 and query 7 from the TCP-H benchmark. Table 4 shows the changes required to implement the Join-Biclique strategy and its ContRand variant for each query. Implementing these strategies proves to be troublesome in Storm, where they require the creation of a grouping, several bolts, and spouts, all of which need to be wired together (and to other components) in the application's topology. In the Skitter implementation, on the other hand, changes are contained to the strategy and workflow languages, as the strategy language is sufficiently expressive to capture the definition of both strategies without the need to change the workflow. Therefore, the only change made to the workflow is the distribution strategy passed to **with:**[6]

**Conclusion** The grouping abstraction provided by Storm suffices to express simple distribution strategies which determine how data is divided over several tasks. However, they are insufficient when expressing more complex distribution strategies. Thus, implementing complex distribution strategies in Storm requires changing the topology, components, and groupings of the application. In contrast, Skitter manages to capture all the distribution logic inside its strategy abstraction.

### 6.3.2 Performance of Distribution Strategies in Skitter

To evaluate if distribution strategies maintain their performance when implemented in Skitter, we run the benchmarks in both Skitter and Storm and compare the results with each other. In this experiment, we mainly verify that Skitter does not change the performance properties of a distribution strategy. Therefore, our discussion focuses mainly on the relative performance of the various distribution strategies compared to one another.

**WordCount** Figure 3 compares the average throughput of the various strategies in Skitter with those obtained by Storm. Both implementations have similar behavior: the D-C, W-C and SG strategies remain largely unaffected by Skew, while the KG and PKG strategies become slower as the skew level ($z$) increases. The rate at which these strategies slow down seems to be similar in both implementations.

**Join** Figure 4 compares the average throughput of the join strategies when used to execute both queries discussed in Section 6.1 on 10 GB of data. The behavior of the strategies is similar in both Storm and Skitter: the Join-Matrix strategy outperforms both variants of Join-Biclique strategy, while the Join-Biclique strategy is significantly outperformed by both the Join-Matrix and Join-Biclique ContRand strategy.

---

[6] 3 and 2 lines are modified, as query 5 uses three joins, while query 7 uses two.

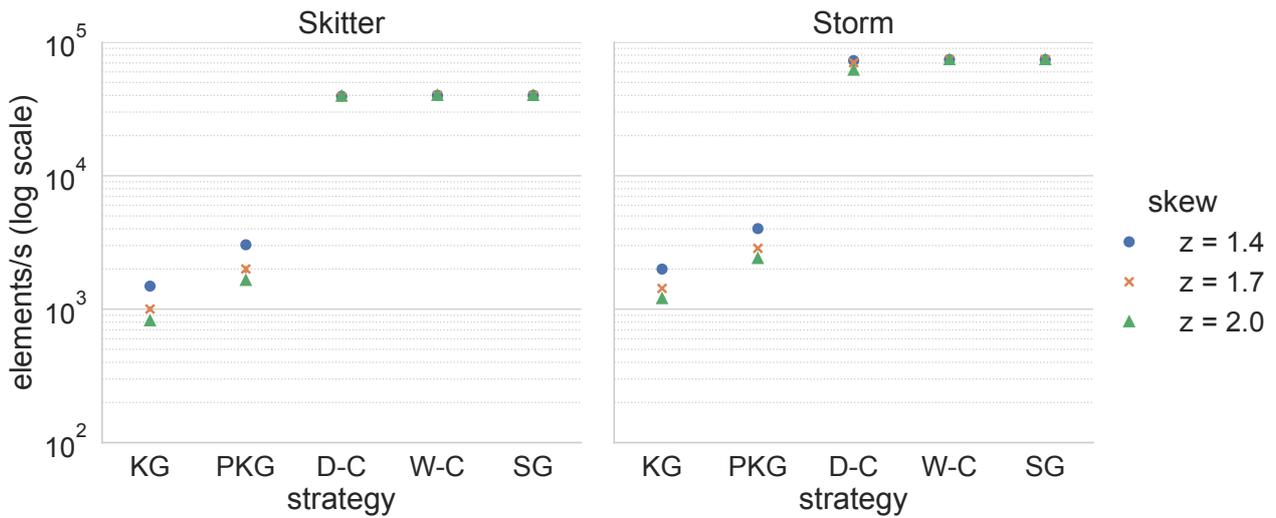

**Figure 3** Comparison of the word count strategies written in Skitter and Storm.

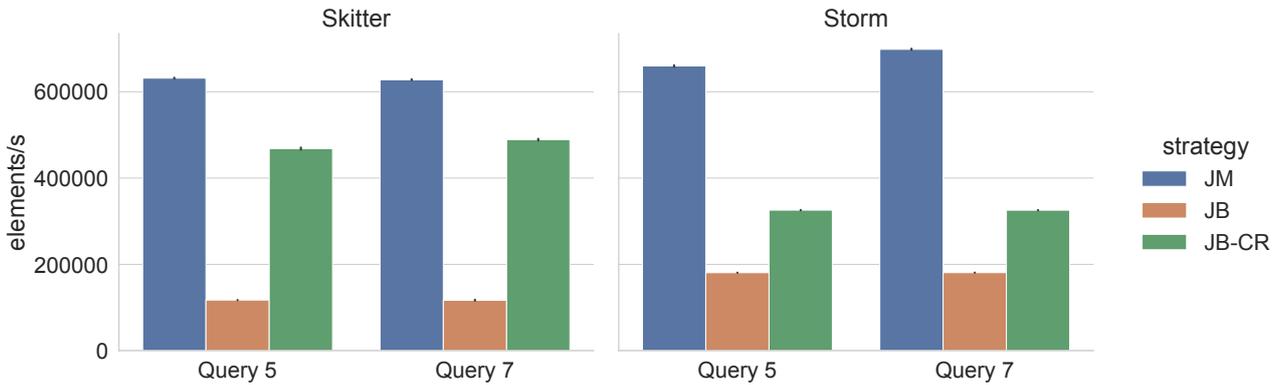

**Figure 4** Comparison of the join strategies for query 5 and query 7 written in Skitter with those written in Storm.

Figure 4 shows that Skitter's Join-Biclique ContRand strategy significantly outperforms its Storm counterpart while the inverse is true for the Join-Biclique strategy. Unfortunately, further investigation does not reveal an immediate cause for this discrepancy; we hypothesize it is related to implementation differences between both DSPFs and by the drastically different virtual machines underpinning them.

**Conclusion** While there are discrepancies between Storm and Skitter, distribution strategies implemented in Skitter showcase the same relative performance as those expressed in Storm. Moreover, applications expressed in Skitter exhibit average throughput rates in the same order of magnitude as those written in Storm. Thus, Skitter enables the modular implementation of distribution strategies without affecting their performance characteristics.

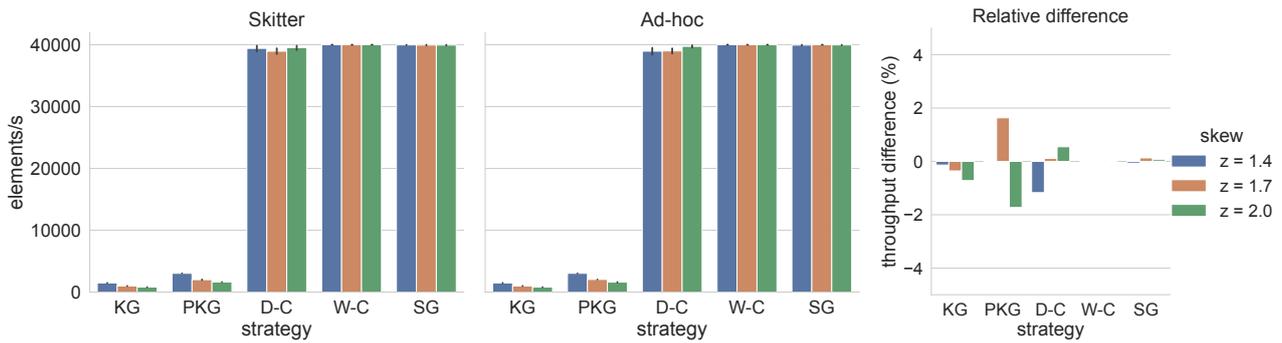

**Figure 5** Average throughput of the word count strategies implemented in Skitter and in Elixir. The right chart shows the difference between both benchmarks.

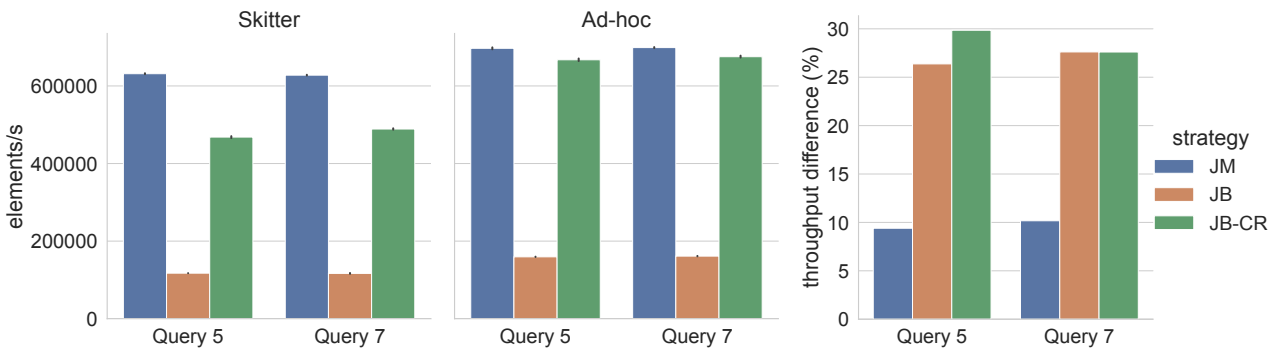

**Figure 6** Average throughput of the join benchmarks implemented in Skitter and in Elixir. The right chart shows the difference between both benchmarks.

### 6.3.3 Overhead Introduced by Skitter

After having found distribution strategies implemented in Skitter exhibit the expected performance characteristics, we measure the computational overhead of Skitter's abstractions. We do this by implementing an ad-hoc version of the experiments described in Section 6.1 in "plain" Elixir, and comparing the performance of this ad-hoc implementation with the Skitter implementation.

**WordCount** Figure 5 compares the average throughput of the Skitter implementation of the word count application with an ad-hoc implementation and shows the difference between both. Both applications exhibit similar average throughput and Skitter outperforms the ad-hoc implementation in some cases, hinting that the abstractions introduced by Skitter do not affect the performance of this benchmark.

**Join** Figure 6 compares the average throughput of the Skitter join application with the ad-hoc implementation. The chart shows that the ad-hoc implementation outperforms Skitter in every case, up to a worst-case difference of 30 %. These differences are related to the meta-information the Skitter runtime system attaches to each data element[7]; this disproportionately affects the Join-Biclique and Join-Biclique ContRand strategies, which move each data element into several data structures before storing them in a join table.

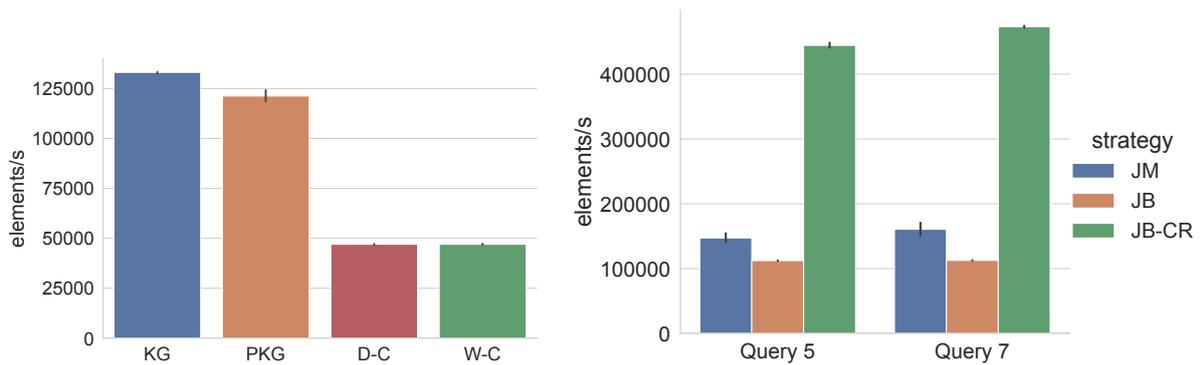

**Figure 7** Average throughput of the WordCount benchmark with key merging and no skew ($z = 0$, left), and of the Join benchmark handling 80GB of data (right).

**Conclusion** In the best case, Skitter applications can exhibit performance characteristics which match or even exceed those of handwritten implementations. However, in certain cases, the additional information tracked by the Skitter runtime system may have an effect on the performance of the application.

### 6.3.4 Effect of Distribution Strategies on Performance

In both of the benchmark applications, a single distribution strategy outperformed all others. In order to show that the performance of an operation can be improved by modifying its distribution strategy, we present a modified version of each application where a different distribution strategy attains the highest throughput.

**WordCount** The left part of Figure 7 shows the average throughput of the WordCount benchmark if the words follow a uniform distribution and if the states of split keys are merged (as described in Section 6.3.1). In this benchmark, the KG strategy outperforms the PKG strategy while also outperforming the D-C and W-C strategies by a significant margin.

**Join** We change the scale of the join benchmark to process 80 GB of data and show the results of this experiment on the right side of Figure 7. In this scenario, the join-matrix strategy's higher memory use causes it to run out of memory on some nodes and use swap space, causing its throughput to drop. The Join-Biclique ContRand strategy is less affected by the increase of received data and outperforms both other strategies in this situation.

---

[7] This information is used to track from which input a data record was received, which enables the implementation of primitives like **port_of** shown on Listing 5, line 8.

**Conclusion** These experiments show that the most performant distribution strategy for an operation is dependent on the properties of the application, and that changing the distribution strategy can improve the performance of a stream processing application.

# 7 Related Work

Meta-programming is a well-known technique to modularise the distribution of programs [16, 25]; moreover, several programming models exist which aim to disentangle performance-critical code from domain-specific code [22]. However, to the best of our knowledge, no work exists which applies these concepts to enable the implementation of distribution strategies in DSPFs. In this section, we discuss how existing DSPFs support distribution strategies and how they compare to Skitter in this regard.

**Apache Storm [23]** Storm is powerful enough for the expression of distribution strategies, but does not provide any explicit abstractions for doing so, which leads to the issues discussed in Section 3. Like Storm, Skitter is sufficiently powerful for the expression of distribution strategies. Unlike Storm, Skitter provides abstractions for the modular expression of these strategies, disentangling them from application-level concerns and making it easier to implement or select the appropriate distribution strategy for a given operation, as we discuss in Section 6.

**High-level DSPFs** Modern DSPFs, such as Spark [3, 27] and Flink [7] disentangle application-level concerns from distribution logic, but do this by creating a fixed set of operations and by hiding the distribution logic of these operations inside the implementation of the framework. The programming model offered by Skitter enables the creation of arbitrary operations and distribution strategies and enables the developer to select the most performant distribution strategy to use for each operation in the application.

Skitter offers a programming model that is a hybrid of the one offered by high-level DSPFs (i.e. it allows workflows to be defined using the operator style, as discussed in Section 4.1), and the one by Storm: Skitter developers can define their own distribution strategies and operations and use them when appropriate.

**Previous versions of Skitter** We discussed an earlier version of Skitter in previous work [21]. This version of Skitter included an early notion of workflows and operations (then called *components*), but did not include any notion of distribution strategies. Instead, a component definition specified several properties which were used by the framework to select one of several predefined distribution strategies, similar to existing high-level DSPFs.

# 8 Limitations and Future Work

Skitter lacks several technical features common to DSPFs used in the industry such as a backpressure mechanism, monitoring infrastructure or a large standard library of built-in operators. Moreover, there is no static enforcement that guarantees that an operation in Skitter implements the callbacks required by a strategy it is coupled with in a workflow. When this is not the case, the application will crash at run-time. We conjecture that there are no scientific reasons why these features cannot be added to Skitter with the required engineering effort, however.

A more interesting limitation of the current incantation of Skitter is its lack of a fault-tolerance mechanism. This is not a fundamental limitation, however. For instance, we have successfully extended Skitter with Storm's data record acknowledgement mechanism without requiring significant changes to Skitter's programming model.[8] However, several different mechanisms for fault-tolerance exists, each with their own strengths and drawbacks [6, 10, 27]. Rather than porting a single fault-tolerance mechanism to Skitter, we believe it would be an interesting research avenue to investigate whether our notion of distribution strategies can be extended to support the expression of failure handling strategies.

Skitter leaves it up to the programmer to decide on which cluster node a worker is created. If the programmer does not specify a node, a random node is selected by the runtime system instead. However, using information such as the current resource usage or hardware characteristics of nodes for scheduling workers can improve the performance of a DSPF [19, 26]. In future work, we would like to investigate if such scheduling techniques can be applied in the context of Skitter.

## 9 Conclusion

In this work, we discuss how the performance of DSPFs is impacted by distribution strategies. Popular DSPFs offer a high-level programming model which does not offer developers the ability to modify the strategies used to distribute the operations in their applications. Alternatively, a model such as Storm allows for the implementation of distribution strategies, but at the cost of tremendous accidental complexity.

To tackle this issue, we present Skitter, a novel programming model which decouples distribution and data processing logic in stream processing applications, allowing developers to choose the most appropriate distribution strategies for their applications. Skitter introduces three main abstractions: operations, workflows and strategies. Operations represent reusable data processing steps; they are composed into a workflow to build a stream processing application. At run-time, operations are distributed over the cluster by distribution strategies, which can be implemented by meta programmers.

We implemented Skitter as a domain-specific language in Elixir and used this implementation to show that our programming model enables the modular implementation of distribution strategies from the state of the art. The results of our evaluation demonstrate that strategies implemented in Skitter exhibit the expected performance characteristics and that the performance of an operation can be improved by modifying its distribution strategy.

---

[8] The most significant change is that strategies need to *anchor* data elements before emitting them, and that sources need to specify how a failed tuple can be handled, similar to how this is done in Storm.

## About the authors

**Mathijs Saey** is a Ph.D candidate at the Software Languages Lab of the Vrije Universiteit Brussel. His research is centered around distributed programming and stream processing applications. Contact him at Mathijs.Saey@vub.be.
https://orcid.org/0000-0002-9481-4181

**Joeri De Koster** is an assistant professor in programming languages and runtimes. His current research is mainly focused on the design, formalisation and implementation of parallel and distributed programming languages. Contact him at Joeri.De.Koster@vub.be
https://orcid.org/0000-0002-2932-8208

**Wolfgang De Meuter** is a professor in programming languages and programming tools. His current research is mainly situated in the field of distributed programming, concurrent programming, reactive programming and Big Data processing. His research methodology varies from more theoretical approaches (e.g. type systems) to building practical frameworks and tools (e.g. crowd-sourcing systems). Contact him at Wolfgang.De.Meuter@vub.be.
https://orcid.org/0000-0002-5229-5627